\documentclass[amsmath,amssymb,preprint,showpacs,nofootinbib]{revtex4-1}
\usepackage{graphicx}
\usepackage{CJK}
\newcommand{\beq}{\begin{equation}}
\newcommand{\eeq}{\end{equation}}
\newcommand{\be}{\begin{eqnarray}}
\newcommand{\ee}{\end{eqnarray}}
\begin{document}
\begin{CJK*}{KS}{}
\title{Transfer matrix algorithm for computing the exact partition function of a square lattice polymer}
\author{Julian Lee}
\email{jul@ssu.ac.kr}
\affiliation{Department of Bioinformatics and Life Science, Soongsil University, Seoul, Korea}
\date{\today}
\begin{abstract}
I develop a transfer matrix algorithm for computing the exact partition function of a square lattice polymer with nearest-neighbor interaction, by extending a previous algorithm for computing  the total number of self-avoiding walks. The computation time scales as $\sim 1.6^N$ with the chain length $N$, in contrast to the explicit enumeration where the scaling is $\sim 2.7^N$. The exact partition function can be obtained faster with the transfer matrix method than with the explicit enumeration, for $N>25$.  The new  results for up to $N=42$ are presented.
\end{abstract}
\maketitle
\end{CJK*}
\section{Introduction}
Polymers play important roles in various fields of science, including biology, where various biopolymers perform crucial functions for life processes. Although properties of heteropolymers such as proteins are most interesting, many important general properties of polymers can be learned from simpler homopolymer models. The simplest toy models for studying such a polymer are lattice models, such as two-dimensional square or three-dimensional cubic lattice polymers~\cite{chan89,chan,flory,degen}. By introducing hydrophobic inter-monomer interactions, a lattice model can be used as a model for a polymer in a dilute solution~\cite{chan89,chan,flory,degen,step,baum,ish,kol,bir,deri,priv,sal,Dup,seno,pool,mei,chang,
gras,bark,nari,chen,zhou,CN,gaud,cara,ponu,VBJ,WL,chen13,lee}. Various quantities such as radius of gyration, end-to-end distance, and specific heat have been calculated for the lattice models.
 
One important advantage of the lattice polymer is that the all possible conformations can be enumerated exactly~\cite{lee,lee2,chen13,chen16}. The exact partition function for lattice polymers up to $N=28$ for cubic lattice and  $N=40$ for square lattice have been computed by a recently developed efficient enumeration algorithm~\cite{chen16}, where $N$ is the number of monomers in the polymer. The most serious obstacle for the explicit enumeration of lattice polymer conformations of longer chain lengths is the fact that the number of conformations and the corresponding computational time grows exponentially with the chain length, as $\sim 2.7^N$\cite{lee2,chen16}. 

In this work, I propose a new transfer matrix approach where the exact partition function     of a square lattice polymer can be computed much faster than using the explicit enumeration for long chains. In the transfer matrix method, instead of generating one conformation at a time, one keeps track of an ensemble of partially built conformations. By throwing away detailed information on partially built conformations and keeping only the essential information required for the computation of the partition function, the transfer matrix method drastically reduce the computational time without sacrificing the exact nature of the computation.  
 The transfer matrix approach has been mostly used for computing the partition function for spin systems~\cite{suzu,skim}, including a simple model of proteins\cite{bru,lee13}.  The most relevant previous work is the transfer matrix method used for the enumeration of  self-avoiding walks(SAWs) on the square lattice~\cite{jenn}. This method is an 
 improvement of earlier methods for enumerating SAWs \cite{jeno,con}, and also an extension of the methods that enumerates the self-avoiding polygons(SAPs) on the square lattice lattice~\cite{jenp,jenpo,ent}. Because a conformation of a lattice polymer is equivalent to a SAW, the {\it total} number of polymer conformations on the square lattice is enumerated by this method. The computation has been performed for up to $N=80$~\cite{jenn}. We generalize this method so that the nearest-neighbor contact between the monomers can be taken into account. By computing the number of conformations for each value of the contact number, the exact partition function can be computed as the function of the temperature.  We find that the computational time scales as $\sim 1.6^N$, in contrast  $\sim 2.7^N$ of  the explicit enumeration. The partition function can be obtained faster with the transfer matrix method than with the explicit enumeration, for $N>25$. All the known results up to $N=40$ can be reproduced within a day with a {\it single} CPU.  The new results for $N=41$ and $N=42$ will also be presented.

\section{Model}
\begin{figure}
\includegraphics[width=.8\textwidth]{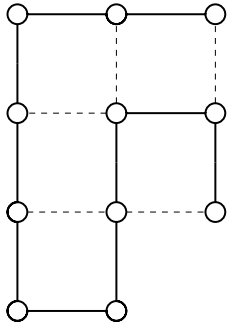}
\caption{An example of a conformation of a square lattice polymer, with $N=11$ and $K=5$. The non-bonded contacts are denoted by dashed lines.}
\label{conf}
\end{figure}

We consider a polymer on a square lattice, where a pair of non-bonded monomers that neighbor each other are regarded as being in contact. The energy value of $-\epsilon$ is associated with each of these nearest-neighbor contacts. Therefore, the energy of a given conformation can be expressed as $E=-\epsilon K$, where $K$ is the number of contacts  formed in the conformation, which will simply be called the contact number from now on. An examples of a square-lattice polymer conformation with $N=11$ and $K=5$ is shown in Fig.\ref{conf}. The partition function is then given by
\begin{equation}
    Z =\sum_{\rm all\ confs} e^{-\beta E} 
    = \sum_{K=0}^{K_{\max}} \Omega (K) e^{\beta \epsilon K},
\label{PF}
\end{equation}
where $\beta \equiv 1/k_B T$, $\Omega (K)$is the number of conformations for a given contact number $K$, also called the density of states. For the polymer on the square lattice, the maximum number of possible contacts $K_{\rm max}$  is given as~\cite{chan89}:
\begin{equation}
    K_{\max}(N) = \left\{
    \begin{array}{ll}
      N-2m & \mathrm{for} ~~ m^2 < N \le m(m+1),\\
      N-2m-1 ~~~ & \mathrm{for} ~~ m(m+1) < N \le (m+1)^2 ,
    \end{array} \right.
\label{Kmax}
\end{equation}
where $m$ is a positive integer and $N$ is the number of monomers forming the polymer chain. It is clear from Eq.(\ref{PF}) that the partition function for any temperature $T$ can be computed once the density of states $\Omega(K)$ is obtained. The purpose of the algorithm developed in the current study is the efficient computation of $\Omega(K)$.  The current model is also called the interacting self-avoiding walk(ISAW) on the square lattice. When $\beta=0$, the partition function of ISAW in Eq.(\ref{PF}) gets reduced to the {\it total} number of SAWs, that has been computed by the transfer matrix approach~\cite{jenn,jeno}.     

\section{Transfer Matrix Method}
\begin{figure}
\includegraphics[width=.8\textwidth] {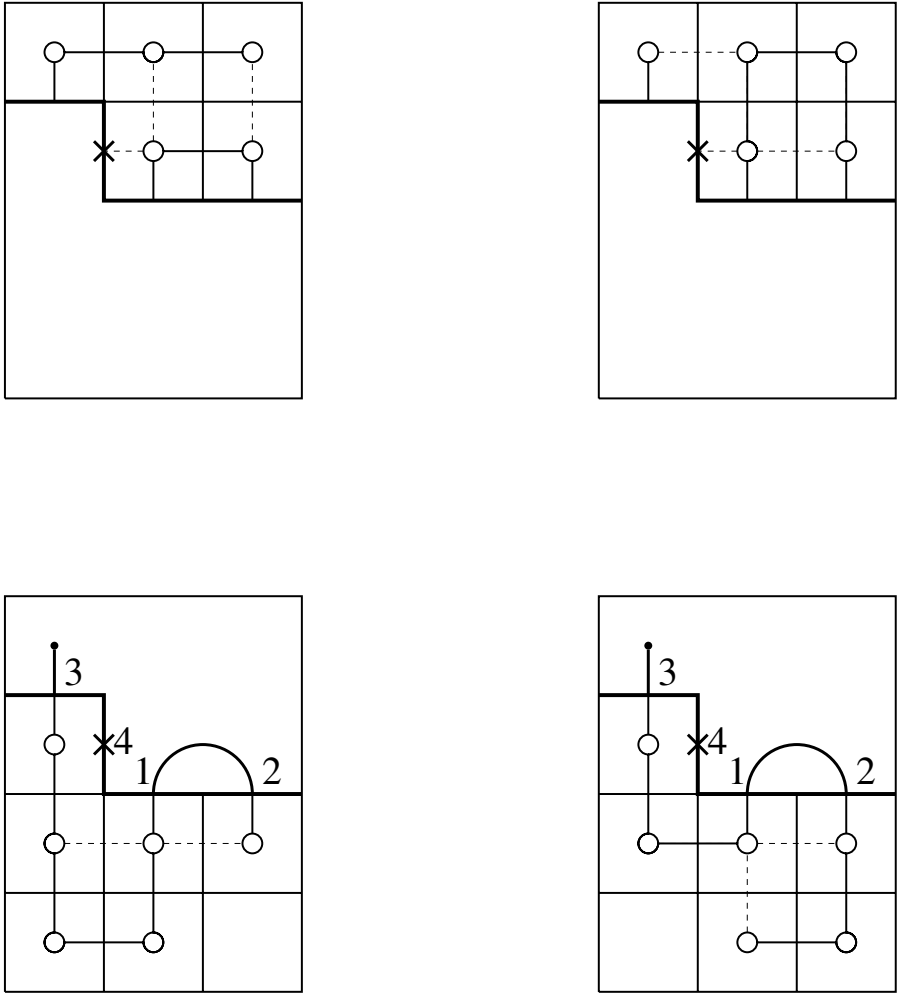}
\caption{An example of building polymer conformations of length $N=11$ spanning the  box of rectangle $3 \times 4$. The current cut-line is shown as the thick line. The cell at $(i,j)=(1,3)$ has been just been completed, and the signature at the cut-line is (3 4 1 2). An  empty edge with a cross on it denotes that the site just beneath the edge is occupied. Two examples of partially built conformation with $n=6$, $k=2$, corresponding to this signature, are also shown below the cut-line at the bottom of the figure.  Two examples of the upper parts of the conformations with $K=5$, generated from this cut-line signature,  are shown at the top.}
\label{cut-line}
\end{figure}

The transfer matrix method developed in the current study is based on a previous method for enumerating the total number of SAWs on the square lattice~\cite{jenn}. First, the conformations are classified according to the rectangular box it spans. For a given box, only the conformations touching all the four walls of the box is enumerated. This idea has also been implemented in a parallel algorithm for explicit enumeration~\cite{lee2}. Let us denote the width and height of the box as $w$ and $h$. It is convenient to visualize the box as consisting of $w \times h$ cells, with each cell enclosing a lattice site~(Fig.\ref{cut-line}). Because the conformation is required to touch all the walls of the box, we get the upper bound for the box size, $w + h  -1 \le N$. In fact, the number of conformations spanning the box with $w + h  -1 = N$ can be obtained by analytic formula, so only the conformations spanning the boxes with $w + h  -1 < N$ need to be enumerated~\cite{lee2}. Because the polymer conformations must fit inside the box, there is also a lower bound $w \times h \ge N$. 

In the transfer matrix method, a cut-line bisecting the lattice is considered, which is moved to build conformations cell by cell~(Fig.\ref{cut-line}).   The main idea of the transfer matrix method is to count the number of partial conformations built up to the cut-line, and use this information to obtain the number of partial conformations when cut-line is moved so that the next cell is incorporated into the lattice space for the partial conformation. This iterative procedure eventually leads to the full density of states when the cut-line reaches the top of the box and all the cells are incorporated. We will take the convention that the initial cut-line is at the bottom of the box, which is moved upward as the algorithm proceeds. For a given row, the cells will be constructed from left to right. Denoting the coordinates of a cell as $(i,j)\ (1 \le i \le  w,\ 1 \le j \le h)$, the cut-line has a kink at the right-hand side of the cell that is included in the partial conformation most recently, and consequently there are $w+1$ edges in the cut-line~(Fig.\ref{cut-line}).

In an earlier version of the algorithm for SAW, the partially built conformations were classified according to their topology of connection to the current cut-line~\cite{jeno}. In the improved newer version, they were classified by the connection topology of the part that is yet to be built, leading to much simpler procedure for pruning out unnecessary conformations~\cite{jenn}. This topological information can be represented by a sequence of digits $s_i\ (i=1, \cdots w+1)$, called the cut-line signature, associated with each edge of the cut-line. In the algorithm so SAW, each digit ranges from 0 to 3, where 0 represents no line crossing the edge, called the empty edge, 1 and 2 the left and the right stems of a loop, and 3 the free end~\cite{jenn}. The new element in the transfer matrix method for ISAW is that in order to keep track of the contact numbers of partial conformations, we need to introduce two types of empty edges depending on whether the site just below the empty edge is occupied of not, denoted by the digits 0 and 4 respectively. Therefore, in the algorithm for ISAW, each digit of the signature ranges from 0 to 4~(Fig.\ref{cut-line}). We will refer to the edge with the digit 4 as being ``charged". At most two digits of the signature can take the value 3, and the number of digits with the values 1 and 2 must be equal~\cite{jenn,jeno}. 

   The number of partial conformations $F(n,k,v)$, called the partial density of states, are recorded for given values of the chain length $n$ and the contact number $k$ of the partial conformations, and a variable  $v$ that records whether the partial conformation has touched the left and the right walls of the box. We will call $n$ and $k$ as the partial chain length and the partial contact number from now on. The variable $v$ is required for the purpose of pruning out unnecessary partial conformations~\cite{jenn,jeno}, as will be explained later. 
The value of $v$  takes four possible values, say from 0 to 3, depending on whether the partial conformation has touched either of the two walls.

In the example of Fig.\ref{cut-line}, polymer conformations of $N=11$ spanning the rectangular box with $w=3$ and $h=4$ are considered.  In the lower part of the figure, the conformations have been built up to the cell at $(i, j) = (1,3)$, and the current cut-line line shown as the thick line. The signature, denoted as (3 4 1 2), describes how the lines coming out of this cut-line will eventually be connected: The line crossing the leftmost edge will become a free end, and the two lines crossing the rightmost edges will join each other to form a loop. The second edge is empty and also charged, because the site beneath the edge, inside the cell at $(1,3)$, is occupied by a monomer. Only the partial conformations consistent with these conditions are being considered. Two examples of such conformations that touch both of the left and right walls, with $n=6$ and $k=2$, are shown at the lower part of the figure. Full conformations generated from a signature must be consistent with the connection topology dictated by the signature . Two examples of the completion of the partial conformations, with $N=11$ and $K=5$, are shown at the upper part of the figure. 
 
\begin{figure}
\includegraphics[width=.8\textwidth]{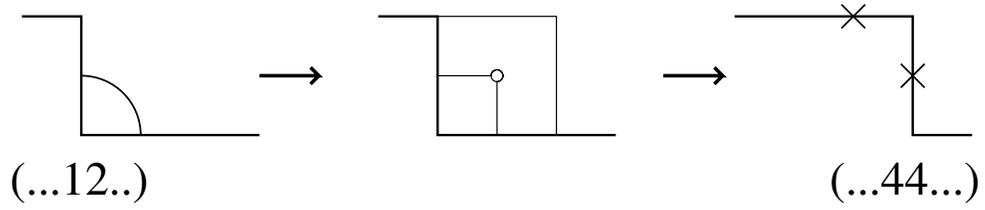}
\caption{An incoming pair of lines (1 2) forms a loop at the new cell to yield a pair of digits $(s_i,s_{i+1})=(4,4)$ in the new signature. The partial chain length $n$ increases by one. The partial contact number $k$ remains unchanged.}
\label{pair}
\end{figure}

\begin{figure}
\includegraphics[width=.8\textwidth]{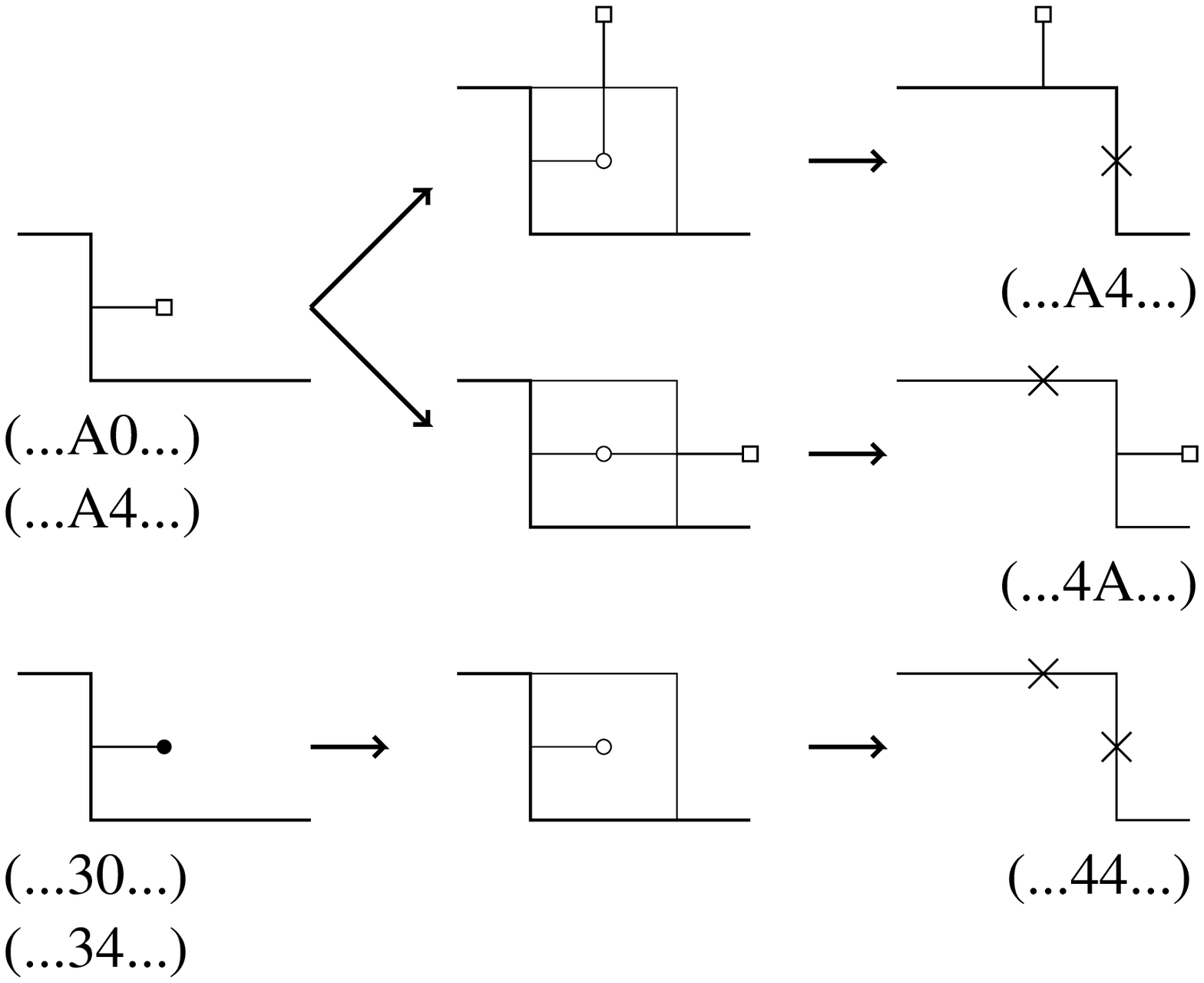}
\caption{A single incoming line from the left edge. The small square at the end of the line denotes either a loop stem or a free end, and the corresponding digit is denoted by $A=1$, $2$, or $3$. The incoming line can continue vertically or horizontally. A free end can also terminate at the new cell. The partial chain length $n$ increases by one. The partial contact number $k$ remains unchanged if the digit associated with the empty edge is 0, or increases by one  if it is 4.} 
\label{hori}
\end{figure}

\begin{figure}
\includegraphics[width=.8\textwidth]{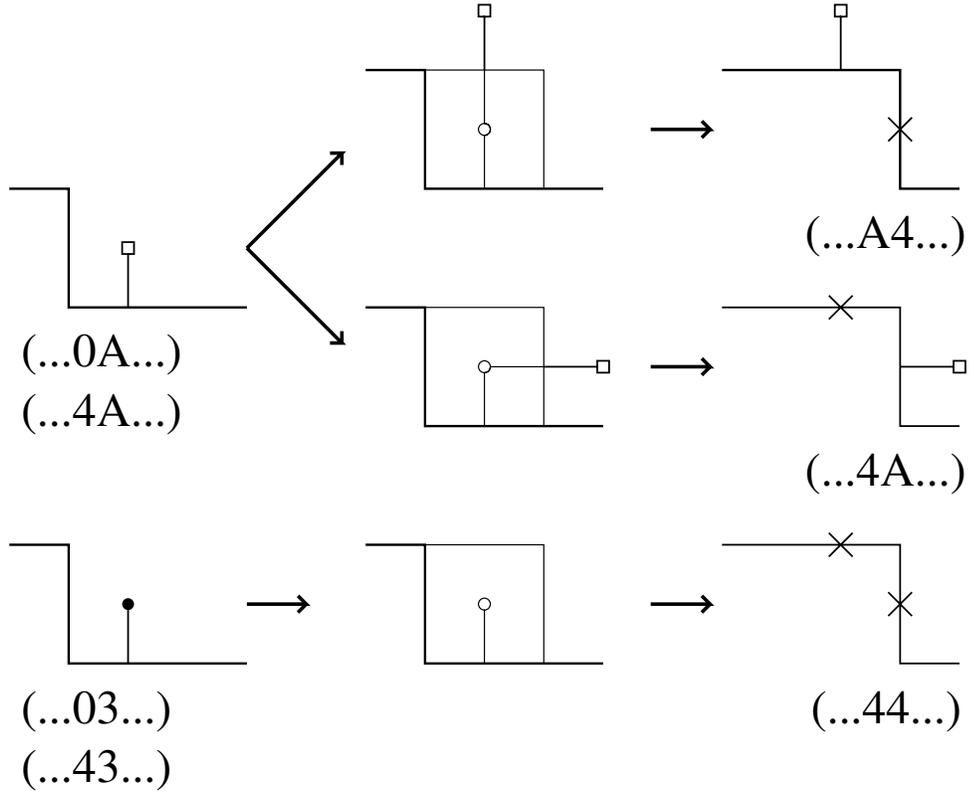}
\caption{A single incoming line from the bottom edge. Square denotes either a loop stem or a free end, and the corresponding digit is denoted by $A=1$,$2$, or $3$. The incoming line can continue vertically or horizontally. A free end can also terminate at the new cell. The partial chain length $n$ increases by one. The partial contact number $k$ remains unchanged if the digit associated with the empty edge is 0, or increases by one  if it is 4.} 
\label{vert}
\end{figure}

\begin{figure}
\includegraphics[width=.8\textwidth]{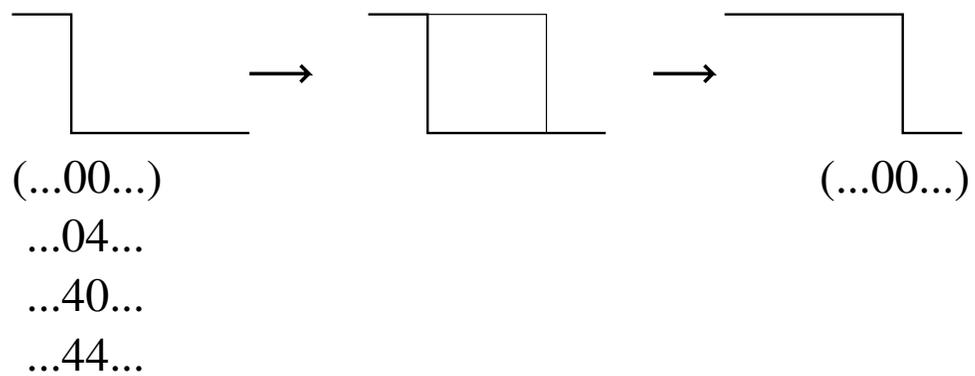}
\caption{No incoming lines. If there is no monomer in the cell, the resulting pair of digits is $(s_i, s_{i+1})=(0, 0)$ in the new cut-line signature. The partial chain length $n$ and the partial contact number $k$ remain unchanged.} 
\label{vacc}
\end{figure}

\begin{figure}
\includegraphics[width=.8\textwidth]{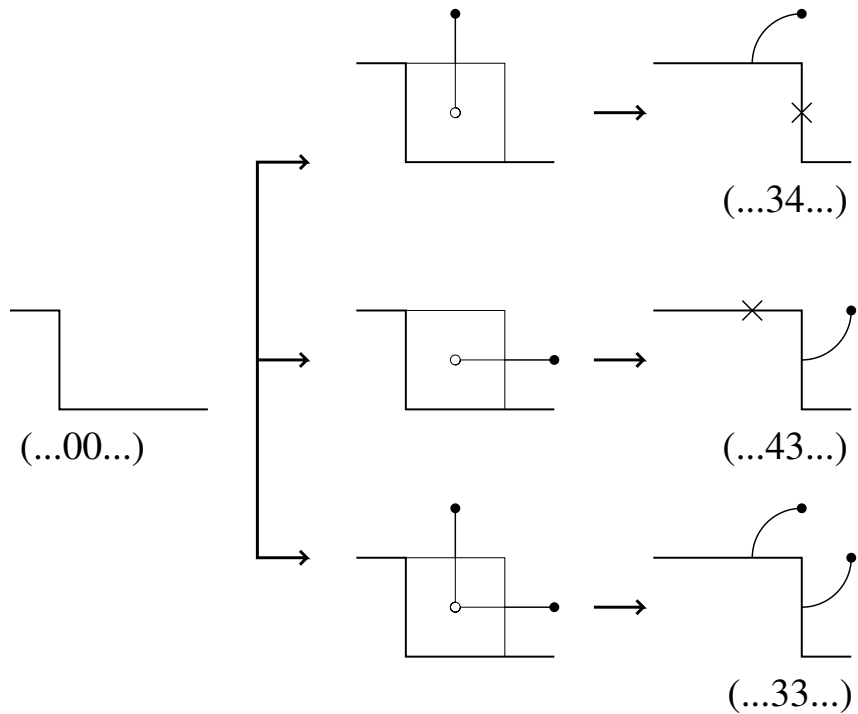}
\caption{No incoming lines, where all the digits of the previous signature are zero. The monomer in the cell is the first one to be encountered. The resulting pair of digits is $(s_i, s_{i+1})=(3,4)$,  $(4,3)$, or $(3,3)$, depending on the number of lines emitted from the monomer and the edge the line is crossing.  The partial chain length $n$ is one and the partial contact number $k$ is zero after any of these updates.} 
\label{crea}
\end{figure}

\begin{figure}
\includegraphics[width=.8\textwidth]{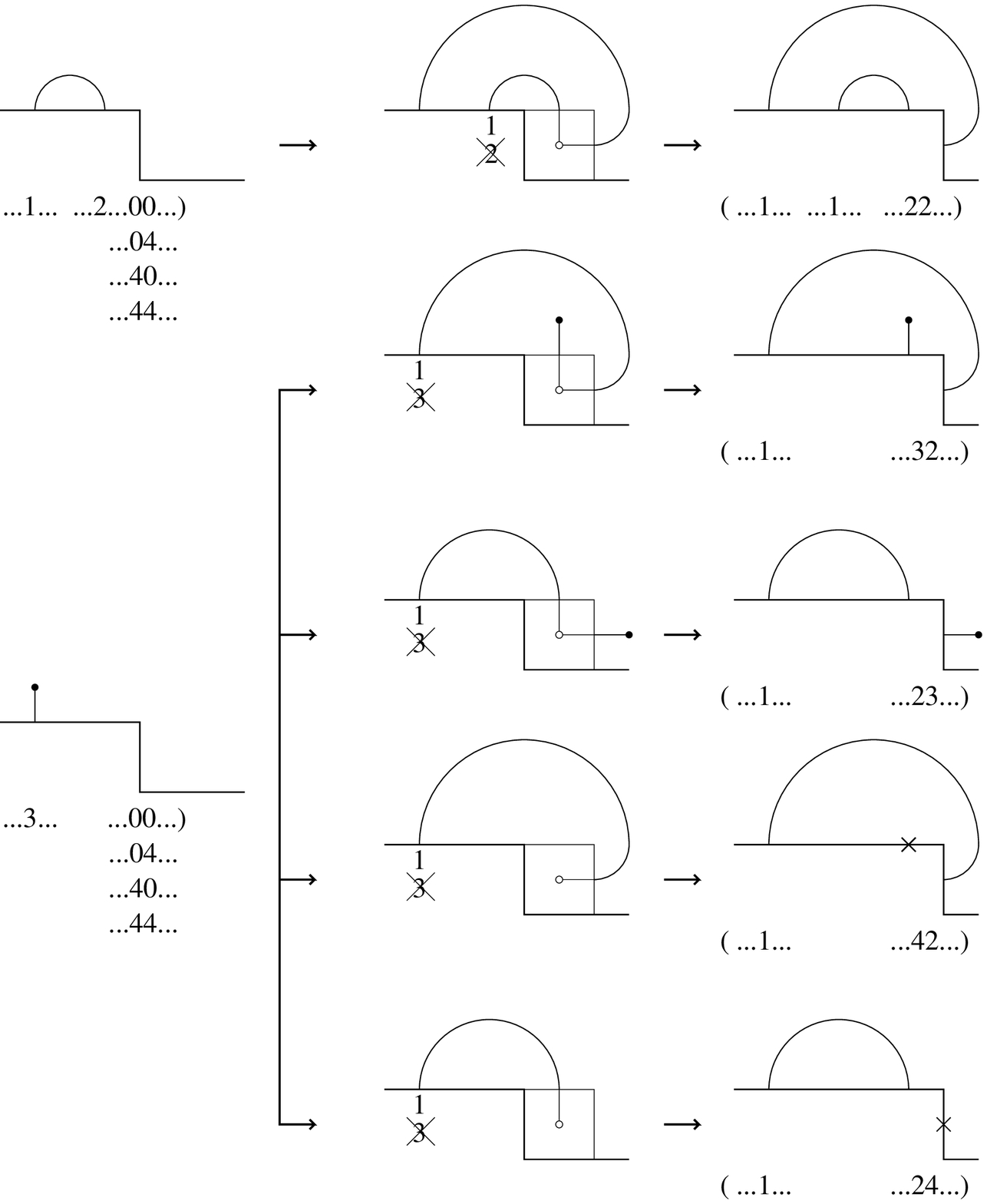}
\caption{No incoming lines. The monomer in the cell comes from the loops or free ends protruding out from the cut-line. The case when the corresponding loop or the free end is located at the left-hand side of the cell is depicted here.  The partial chain length $n$ increases by one. The partial contact number $k$ remains unchanged if both of the digits associated with the empty edges are 0, increases by one if one of them is 4, or by two if both of them are 4.} 
\label{left}
\end{figure}

\begin{figure}
\includegraphics[width=.8\textwidth]{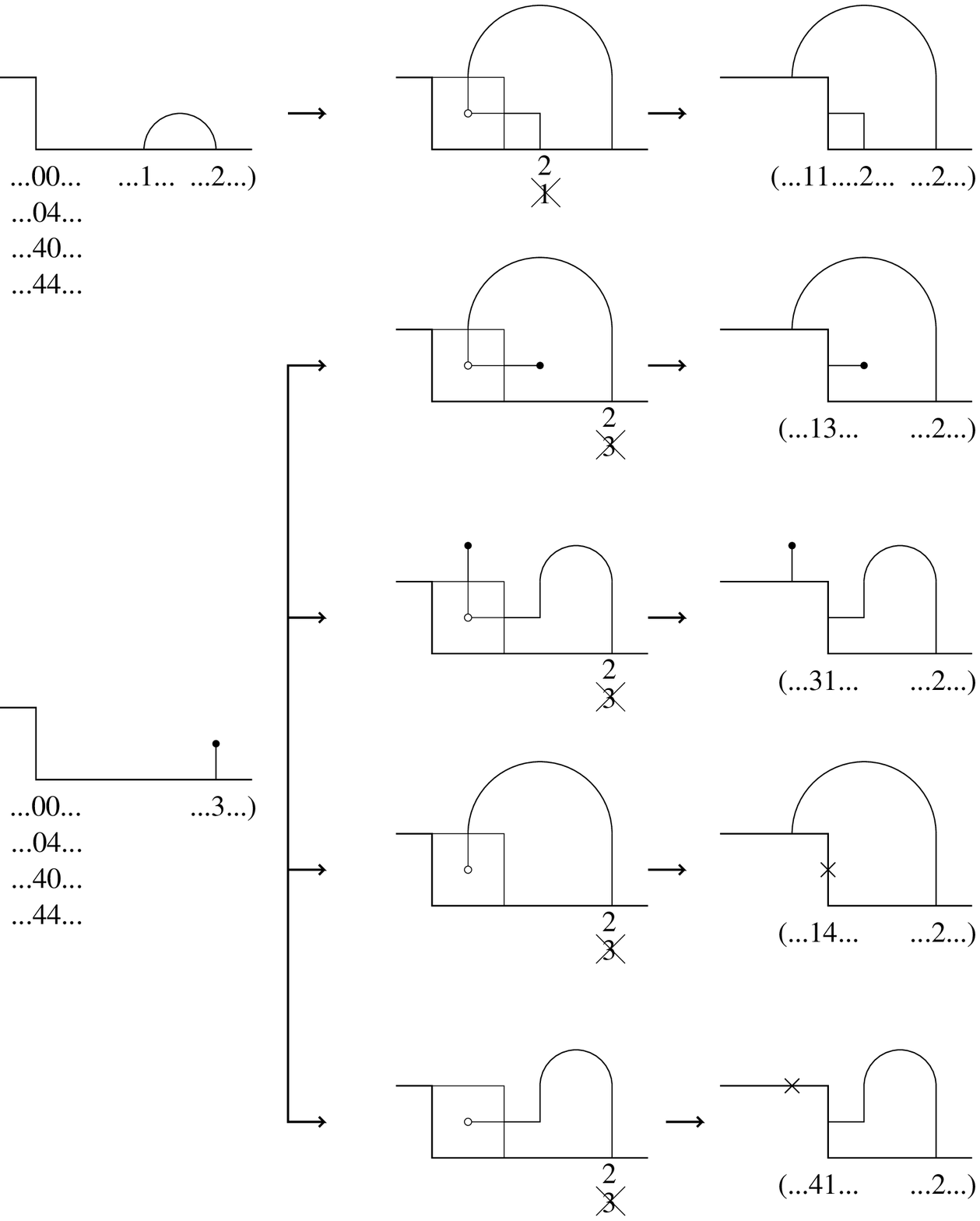}
\caption{No incoming lines. The monomer in the cell comes from the loops or free ends protruding out from the cut-line. The case when the corresponding loop or the free end is located at the right-hand side of the cell is depicted here.  The partial chain length $n$ increases by one. The partial contact number $k$ remains unchanged if both of the digits associated with the empty edges are 0, increases by one if one of them is 4, or by two if both of them are 4.} 
\label{right}
\end{figure}

\begin{figure}
\includegraphics[width=.8\textwidth]{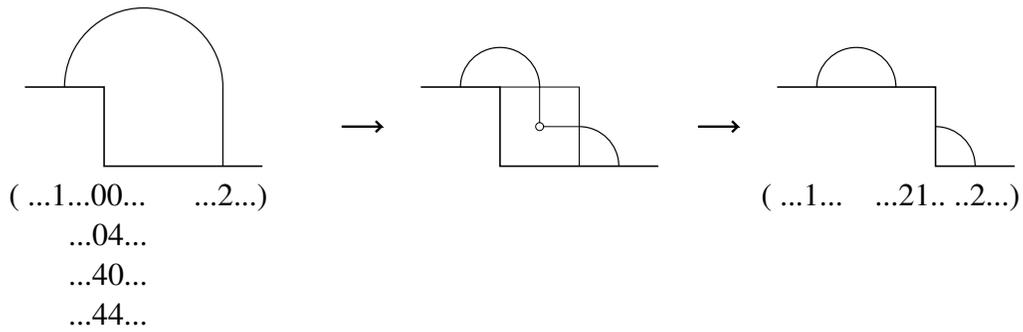}
\caption{ Joining of the loop whose stems are located at the left and the right-hand side of the cell.  The partial chain length $n$ increases by one.  The partial contact number $k$ remains unchanged if both of the digits associated with the empty edges are 0, increases by one if one of them is 4, or by two if both of them are 4.} 
\label{mid}
\end{figure}

The cell-by-cell construction of an ensemble of partial conformations can be described in general terms by the update rules of a cell.  This is exactly the same as that in the previous method for SAW~\cite{jenn}, except that we distinguish empty edge according to whether the site directly beneath is occupied by a monomer, in order to keep track of contact numbers. At the moment when a new cell is being added to the lattice space for the partial conformation, we will call the left and the bottom edges of the cell, which used to be the parts of the previous cut-line, as the incoming edges. Similarly, the right and the top edges, which will become the parts of the new cut-line, will be called the outgoing edges. We will also call the polymer bonds that cross the incoming and outgoing edges as the incoming and outgoing lines, respectively~(Figs.\ref{pair}-\ref{mid}).

The simplest case is when both of the incoming edges are occupied, as shown in Fig.\ref{pair}. The lines crossing the left and bottom and edges must be 1 and the 2 be consistent, In this case, they join at the monomer in this cell and  become part of the partial conformation below the new cut-line. The numbers $(s_i\ s_{i+1}) = (1\ 2)$ encoding this loop in the previous cut-line signature turns into charged empty edges $(s_i\ s_{i+1}) = (4\ 4)$ in the new signature. Any other incoming pair of lines leads to an inconsistency and is not allowed. In fact, the update rule for a single incoming line is such that incoming pairs of lines other than (1 2) never appear, as elaborated below. The partial chain length $n$ increases by one and the partial contact number $k$ remains unchanged, after this update.

When there is a horizontal single incoming line of the form $(s_i\ s_{i+1}) = (A\ 0)$ or $(A\ 4)$, where $A=1,2$, or $3$, this line can continue through either the right or the top edge. The vertical and the horizontal continuation lead to the pair $(s_i\ s_{i+1}) = (A\ 4)$ and $(4\ A)$ in the new cut-line signature, respectively, as shown at the upper part of the figure \ref{hori}. The horizontal continuation is allowed only if $s_{i+2}=0$, or $A=1$ and $s_{i+2}=2$. This ensures that the inconsistent pair of incoming lines at the next cell does not appear. For the case when the incoming line is a free end, $A=3$, there is another possibility that the line terminates in the new cell. This leads to $(s_i\ s_{i+1}) = (4\ 4)$ in the new cut-line signature, as shown at the lower part of the figure. Similar update rule applies for the vertical input line~(Fig.\ref{vert}). Again, the horizontal continuation is allowed only if $s_{i+2}=0$, or $A=1$ and $s_{i+2}=2$. The partial chain length $n$ increases by one for all of these updates. The partial contact number $k$ increases by one or remains unchanged depending whether the empty edge is charged or not.

When there are no incoming lines, the simplest case is the one where there is no monomer in the new cell, with the resulting pair of digits given as $(s_i\ s_{i+1})=(0\ 0)$ in the new signature~(Fig.\ref{vacc}). This is the only update where the partial chain length $n$ remains the same as the previous state. Consequently, $k$ also remains unchanged.

The case with a monomer in the new cell is more complicated. This could be the first time we encounter a monomer of the partial conformation, in which case there may be one line or two lines connected to this monomer going vertically upward or horizontally right, leading to one free end or two free ends~(Fig.\ref{crea}). Because we are considering a single connected chain, and the conformation is required to touch the bottom wall of the box, this kind of update is allowed only at the first row, and only for $n=0$. As a consequence of $n=0$, $s_k=0$ for all $k$ in the previous signature. The partial chain length and the partial contact number are $n=1$ and $k=0$ after any of these updates.

If the monomer in the current cell is not the first monomer we encountered, then it is a part of the loop or free end protruding out from the cut-line, coming back into the cut line through the top or the right edge. The case where the corresponding loop or the free end is located at the left of the new cell is depicted in Fig.\ref{left}. In contrast to the updates rules considered so far, the update rule here is non-local in that a digit far away from the edges of the current cell is also modified. Consider the case of loop joining, shown at the top of the figure. Not only the pair $(s_i, s_{i+1})$ of the cut-line signature gets updated to $(2\ 2)$, but digit associate with the right stem of the loop that has been joined, changes from $s_k=2$ to $s_k=1$. Note that this ensures the balance of $1$ and $2$ in the new cut-line signature. Similar non-locality appears when a terminal is joined, as shown for four cases in the lower part of the figure. Again, not only  $(s_i, s_{i+1})$ gets modified, but the digit associated with the edge where the free end is coming out, changes from $s_k=3$  to $s_k=1$ in the new cut-line signature, because now the protruding line forms a loop. The new pair of digits at the edges of the current cell is $(2\ 3)$, $(3\ 2)$, $(2\ 4)$, or $(4\ 2)$, depending on whether this line is entering through the right or the top edge, and whether the free end protrudes out of the current cell or terminates there, as shown in the figure. Similar update rules exist for the case when a loop or a free end at the right side of the cell passes through or terminates at the current cell(Fig.\ref{right}). Finally, the loop whose stems are at the left and the right-hand side of the current cell can be joined. In this case, only the digits associated with the current cell get modified, to $(s_i, s_{i+1})=(2\ 1)$~(Fig.\ref{mid}). The most difficult and time-consuming part of this update procedure is to find all the loops and free ends that can be joined in this manner. The method is exactly the same as that for SAW~\cite{jenn}. After any of these updates, the partial chain length $n$ increases by one. The partial contact number $k$ increases by the number of charged edges.
   
We note that conformations of ISAW related by discrete rotations and reflections contribute the same amount to the partition function. It is rather straightforward to remove this symmetry in the case of explicit enumeration~\cite{lee2,chen16}, but this is not the case for the transfer matrix computation. 
Because the discrete rotational and reflectional  symmetry is eight-fold~\cite{lee2},  only a four-fold symmetry remains for the conformations spanning a non-square box, if we consider only the boxes with $w < h$.  We also note that in the transfer matrix method, only undirected conformations of lattice polymers are generated, whereas we want the number of conformations for directed polymers. Therefore, we have to multiply the results of transfer matrix computation  by two to distinguish two directions. Therefore, for the number of conformations spanning a non-square box, we have to divide the density of states by two to obtain the symmetry-reduced density of state $\omega(K)$ for directed polymers. In the case of a square box where the rotational and reflection symmetry is eight-fold, we have to divide the result by four to obtain $\omega(K)$.

\section{Pruning}
In the cell-by-cell building procedure described above, many unnecessary cut-line signatures appear, which cannot lead to legitimate full conformations. By removing these unnecessary states as early as possible, the computational time can be drastically reduced. 
In fact, the reason that the future connection topology was used for the cut-line signature in the new version of the SAW transfer algorithm, rather than the past connection topology as was used in the older version, is because the specification of the future connection topology simplifies the pruning procedure considerably~\cite{jenn}. One can compute the minimal number of monomers required for making connections specified by the cut-line signature. Additional monomers may be needed to touch the top of the box. Also, if the value of $v$ indicates that the partial conformation does not touch either of the left or the right walls of the $w \times h$ box, then additional monomers  may be required in order for the remaining part of the conformation to touch the corresponding wall. If the number of unused monomers, $N-n$, is less than the minimal required number of monomers, than the current combination of $\{s_i\}$, $v$, and $n$ is pruned out and prevented from generating future conformations. The method is exactly the same as that in the case of the enumeration of SAWs~\cite{jenn}. 

In addition, if the minimal required number of monomers is less than the remaining volume of the box, or the height of the minimal-length configuration exceeds the remaining height of the box, the current state is pruned out.


\section{Computational Time}
At each cell of the lattice, the transfer matrix generates new cut-line signatures from the old ones. From the update rules, it is clear that the number of new cut-line signatures generated from a given old cut-line signature is of order one, so the computational time for processing a cell at $(i, j)$ will be proportional to the number of cut-line signatures prior to the processing the cell, denoted as $N_s(i,j; N,w,h)$, and the total computational time $t(N,w,h)$ for computing the density of states of the polymer conformation of length $N$ spanning the box of width $w$ and height $h$ will be proportional to
\begin{equation}
t(N,w,h) \propto \sum_{j=1}^h  \sum_{i=1}^{w+1} N_s(i,j; N,w,h). 
\end{equation}
Because only the numbers 0,1,2,3 or 4 can appear at each digit of the cut-line signatures, $N_s(i,j; N,w,h) < 5^{w+1}$. This is a strict inequality because actually there are many constraints such as the fact that the terminals appear at most twice, the loop stems have to be balanced, etc. This bound leads to the upper bound for the computational time:
\begin{equation}
t (N,w,h) < (w+1) h 5^{w+1} \times {\rm const}. \label{rb}
\end{equation}
Because of the symmetry of the problem, we may restrict the computation to the boxes with $w \le h$ or $w \ge h$ both of which yield  the same result, but the restriction to the boxes with $w \le h$ will lead to less number of intermediate cut-line signatures and hence less computational time. From the conditions $w \le h$ and $w + h  \le N$, we get the upper bound for $w$, $w \le N/2$. Therefore, the computational time $T(N)$ for the chain length $N$ satisfies the inequality:
\begin{eqnarray}
T(N) &=& \sum_{w,h} t(N,w,h) <\sum_{1 \le w \le N/2}  N (N/2+1) 5^{N/2} \times {\rm const}  = N^2 (N+2) 5^{N/2} \times {\rm const} \nonumber\\
&=& N (N+2) 2.24^N \times {\rm const}. \label{rb2}
\end{eqnarray}
Asymptotically, this upper bound is better than $2.7^N$ of explicit enumeration, showing that in the limit of $N \to \infty$, the transfer matrix is superior to the explicit enumeration in terms of computational time. Of course this asymptotic result may be of little use if the overall multiplicative constant in Eq.(\ref{rb2}) is too large.  Considering the worst scenario, it might be the case that the transfer matrix is slower for short chain lengths, and the length where the computational time of the transfer matrix method becomes comparable to that of the explicit enumeration, is much longer than the range of lengths accessible to present day computers. One fortunate thing is that the upper bound in  Eq.(\ref{rb2}) is very loose, and there is a large gap between actual $T(N)$ and this upper bound. It is crucial to reduce $T(N)$ further by pruning out unnecessary cut-line signatures at early stages, as explained in the previous section.

\begin{figure}
\includegraphics[width=.8\textwidth]{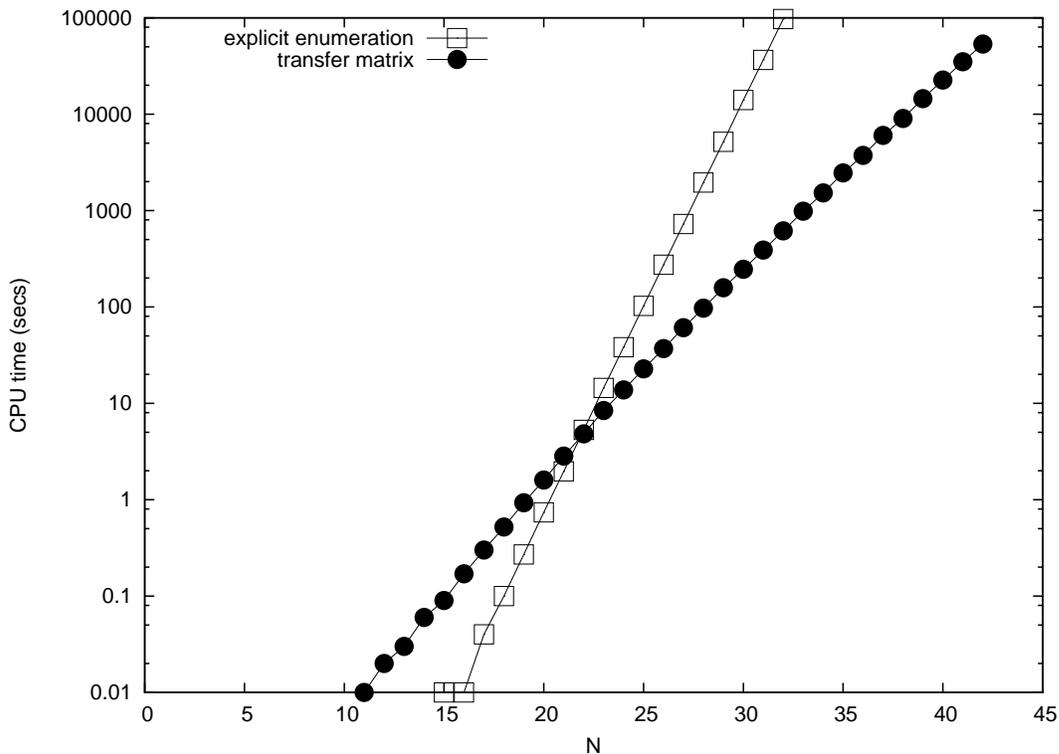}
\caption{ The computational times of the transfer matrix and the explicit enumeration compared.} 
\label{time}
\end{figure}

\begin{figure}
\includegraphics[width=.8\textwidth]{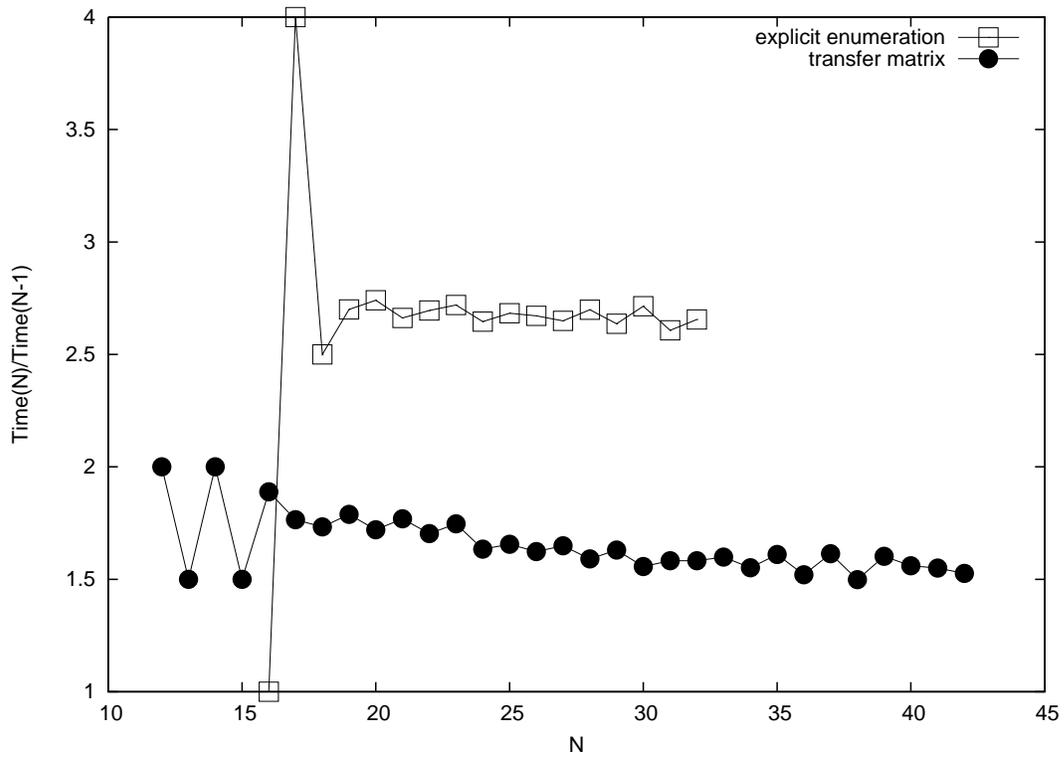}
\caption{The ratio of the computational times for the chain length $N$, to that for $N-1$, shown for both the transfer matrix computation and explicit enumeration.} 
\label{ratio}
\end{figure}

The actual computational time of the transfer matrix, as well as that of the explicit enumeration, are shown in Fig.\ref{time} as the functions of chain length, up to $N=32$ for the explicit enumeration and up to $N=42$ for the transfer matrix. The program was written in C language and compiled with -O3 option, and run on {\it single} Intel i3-3220 CPU. We find that the computational time of the explicit enumeration follows the same scaling as that for the total number of conformations, and scales as $\sim 2.7^N$.  On the other hand, the computational time for the transfer matrix scales as $\sim 1.6^N$. This is more clearly seen in Fig.\ref{ratio} where the ratio of computational time between the adjacent chain lengths are shown as the function of the chain length for each method. The scaling of   $\sim 1.6^N$ for the computational time of the transfer matrix is a rather      conservative estimate: The ratio of the computational time for $N=42$ to that for $N=41$ is in fact about $1.52$, suggesting that the ratio may approach 1.5 asymptotically~(Fig.\ref{ratio}).


From these results, we see that the transfer matrix computation is faster than the explicit enumeration for $N > 21$. We note that the recent efficient implementation  
of explicit enumeration has increased the computational speed considerably~\cite{chen16}. The improvement relevant for the serial computation in a single CPU is the one-step generation of the last two monomers in the chain. This will correspond to the decrease of the  effective chain length by two in terms of the computational time, resulting in the rightward horizontal shift of the graph for the explicit enumeration in Fig.~\ref{time}. Taking this into account, we may safely say that the transfer matrix is faster than the explicit enumeration for $N > 25$. In fact, the computation time for chain length 42 using the transfer matrix method takes only 15 hours on the {\it single}  CPU, whereas the explicit enumeration is expected to take about six years even if we generate the last two monomers at one step.

\section{Memory requirement}
In the case of the explicit enumeration, only the occupation status of the lattice sites are to be recorded at any moment, so the demand for the memory is virtually negligible. On the other hand, the transfer matrix requires a considerable amount of memory, because the intermediate cut-line signatures must be stored at each step. At each step, we generate the new partial densities $F_{\rm new} ( s_1, \cdots s_{w+1},v,n,k)$ from the old ones $F_{\rm old} ( s_1, \cdots s_{w+1},v,n,k)$. After the generation of $F_{\rm new}$, $F_{\rm old}$ is no more needed, so its memory space can be recycled and be used as that for $F_{\rm new}$ at the next step. Therefore, we only need the memory storage for  $F_{\rm new}$ and $F_{\rm old}$.

\begin{figure}
\includegraphics[width=.8\textwidth]{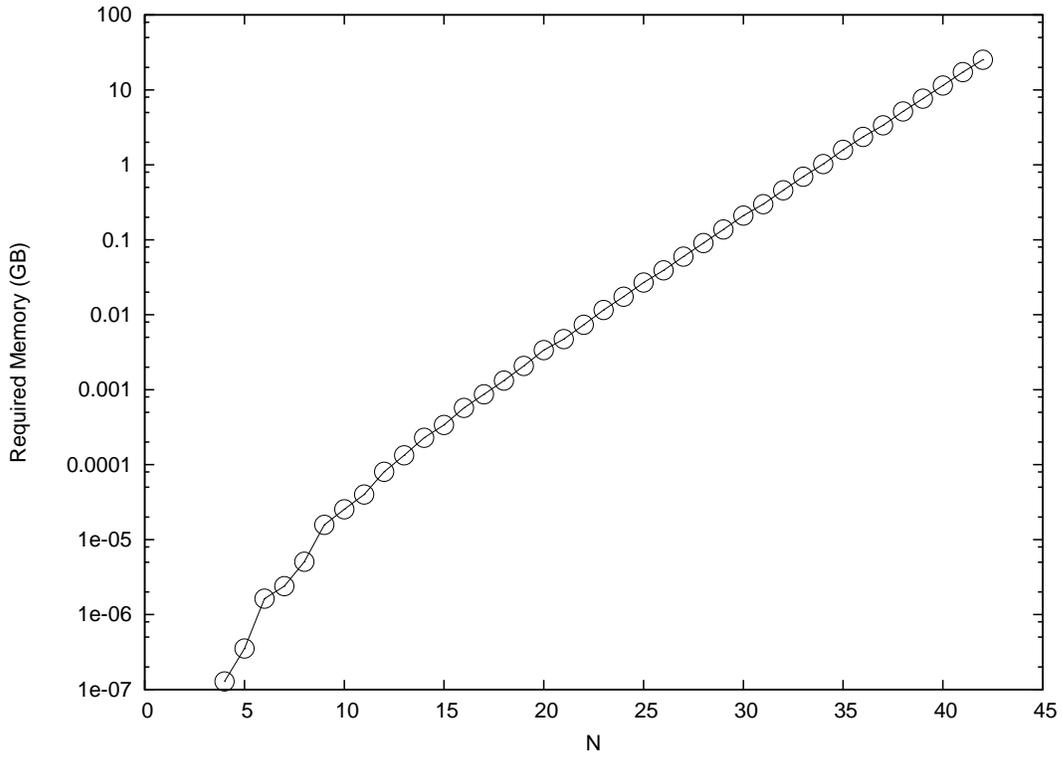}
\caption{The memory requirement for the partial density of states, as the function of the  chain length $N$.} 
\label{memory}
\end{figure}

\begin{figure}
\includegraphics[width=.8\textwidth]{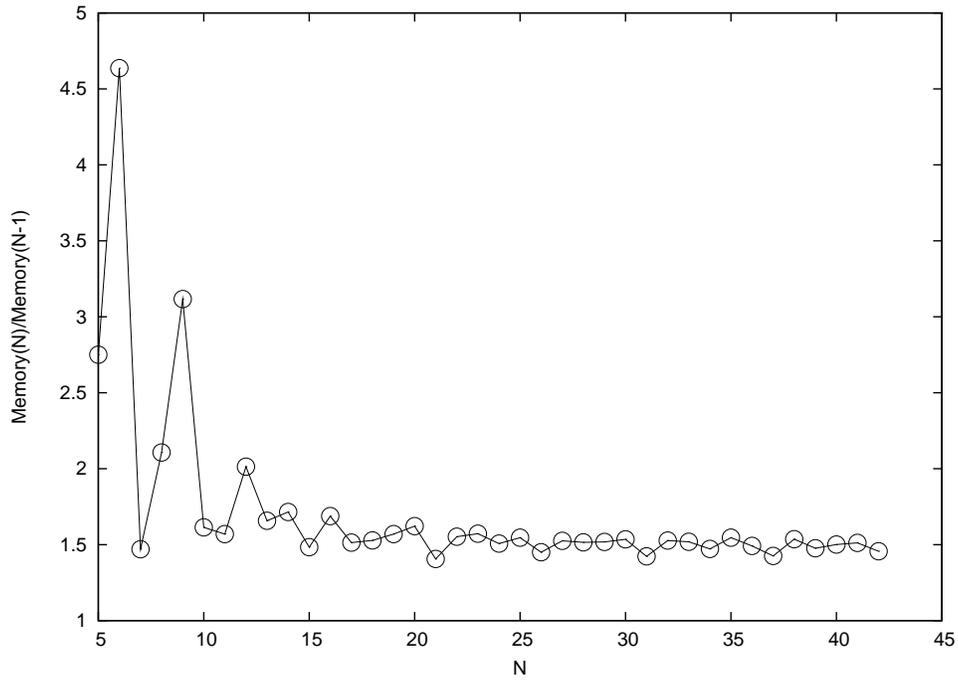}
\caption{The ratio of memory size for the chain length $N$, to that for $N-1$.} 
\label{rmem}
\end{figure}

We cannot allocate the memory space of reasonable size for $F ( s_1, \cdots s_{w+1},v,n,k)$ using standard dynamic array, especially because we cannot predetermine a reasonable value of upper bound for the number of possible signatures $(s_1, \cdots s_{w+1})$.  Therefore, we used the  red-black tree~\cite{RB}, a data structure whose size increases as new items are inserted, which also support fast retrievals and insertions of items, to store the combinations of $(s_1, \cdots s_{w+1},v)$. Because $0 \le n \le N$ and $0 \le k \le K_{\rm max} (N)$, the density of states may be stored as an array of size $(N+1) \times ( K_{\rm max} (N)+1 ) $ for a given combination of $(s_1, \cdots s_{w+1},v)$. However, in order to save memory space further, we stored the partial density of states only for the values of $n$ with non-zero number of conformations, using the linked list. The partial density of states $F ( s_1, \cdots s_{w+1},v,n,k)$ for a given combination of $(s_1, \cdots s_{w+1},v,n)$ was then stored as an array of size $K_{\rm max} (N) +1 $.  Because $F ( s_1, \cdots s_{w+1},v,n,k)$ is stored as an eight-byte unsigned long integer, the memory requirement $N_{\rm mem}$ in bytes is
\begin{equation}
N_{\rm mem}=16N_c^{\rm max} (K_{\rm max} (N) + 1), \label{mem}
\end{equation}
neglecting the space for other variables. Here, $N_c^{\rm max}$ denotes the maximum number of combinations $(s_1, \cdots s_{w+1},v,n)$ encountered during the progression of the algorithm. The extra factor of two comes from the fact that the memory space for  both $F_{\rm new}$ and $F_{\rm old}$ is required\footnote{Of course the numbers of the old  and the new signatures are different at any moment in general, so the memory space needed would be slightly less than that given in Eq.(\ref{mem}), but we neglect this small difference.}. The values of $(K_{\rm max}+1)$, $N_c^{\rm max}$, $(K_{\rm max}+1) \cdot N_c^{\rm max}$ and $N_{\rm mem}$ are given in table 1 as the functions of $N$, and the graph for  $N_{\rm mem}$ is shown in Fig.\ref{memory}. The memory asymptotically scales as $\sim 1.5^N$, as can be seen from the graph for the ratio of memory space for chain length $N$ to that for $N-1$~(Fig.\ref{rmem}). The chain length of up to $N=46$ seems feasible with 128 GB memory. Memory space can be saved further by storing only the non-zero partial density of states for a given combination of  $(s_1, \cdots s_{w+1},v,n)$, instead of using an array of fixed size $K_{\rm max}(N) +1$.
We have not implemented this additional flexibility at this stage, because it will make the code unnecessarily complicated.

\section{New results}
With the transfer matrix, all the known results for up to $N=40$ could be reproduced~\cite{lee2,chen16,38}, and the new results for $N=41$ and $N=42$ could be obtained. The symmetry reduced densities of states $\omega(K)$ are shown for $N=41$ and $N=42$ in Table 2. The correctness of the result can be also cross checked against the {\rm total} number of SAWs, $\sum_K \Omega(K)$, previously obtained using a transfer matrix algorithm for up to $N=80$~\cite{jenn,jeno,Jenw}.

It is straightforward to compute the exact partition function from the density of states using the formula Eq.(\ref{PF}), from which various physical quantities can be obtained. One example of such a quantity is the specific heat per monomer:
\begin{eqnarray}
\frac{C}{k_B N }  &=& \frac{1}{N k_B} \frac{\partial \langle E \rangle}{\partial  T} = \frac{1}{N k_B^2 T^2}\left(  \frac{\partial^2 \ln Z}{\partial \beta^2} \right)\nonumber\\
&=& \frac{1}{Nk_B^2 T^2}\left[ \frac{\sum_q q^2 \Omega(q) e^{\beta q \epsilon}}{\sum_p \Omega(p) e^{\beta p \epsilon}} - \left(\frac{\sum_q q \Omega(q) e^{\beta q \epsilon}}{\sum_p \Omega(p) e^{\beta p \epsilon}}\right)^2 \right].
\end{eqnarray}

\begin{figure}
\includegraphics[width=.8\textwidth]{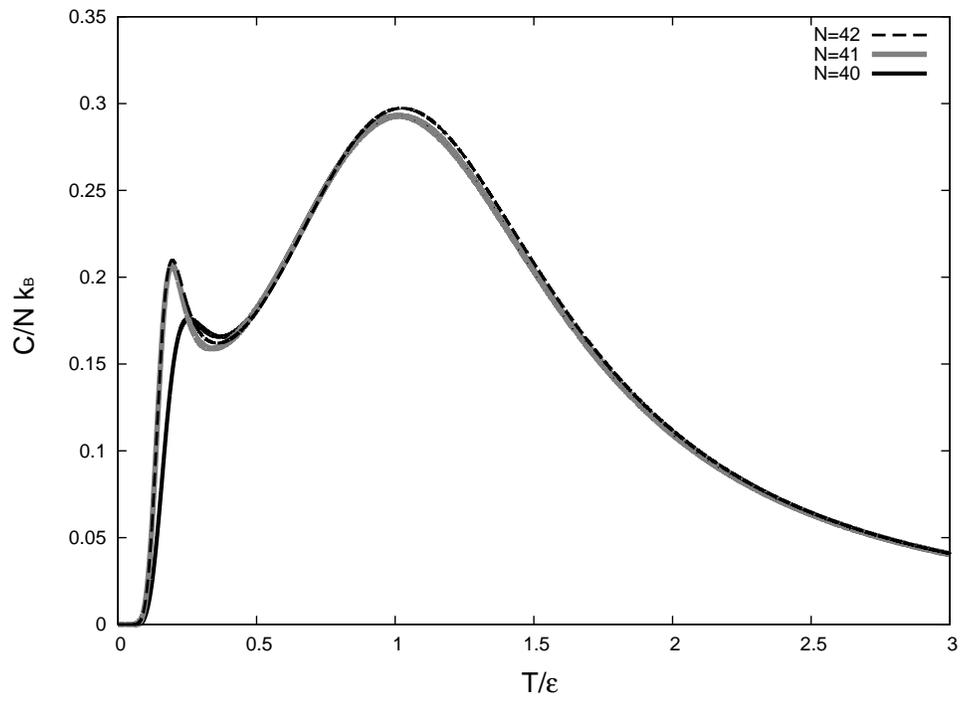}
\caption{Specific heat per monomer, $C/N k_B$, as function of $T/\epsilon$, for $N=40$, $41$, $42$.} 
\label{sh}
\end{figure}

The specific heat for $40 \le N \le 42$ are shown in Fig.\ref{sh} as the functions of temperature $T/\epsilon$. The two peaks at $T/\epsilon \simeq 1.0$  and $T/\epsilon \simeq 0.2$ correspond to collapse and freezing transitions~\cite{VBJ,WL,chen13}. As can be seen in the figure, in contrast to the peak for the collapse transition that change smoothly with $N$,  the peak for the freezing transition becomes especially prominent as $N$ approaches 42. This is because $42=6 \times 7$ is the magic number where the ground states have a special form and their numbers are smaller than for neighboring values of $N$~\cite{VBJ,WL,magic}. There is a possibility that this transition is only a finite-size effect and does not exist in the infinite size limit. The study of polymers with longer chain sizes may shed more light on this issue.

\section{Discussion}
In this work, I developed a transfer matrix method for computing the exact partition function of ISAW model on the square lattice, by extending the previous algorithm for the enumeration of the total number of SAW conformations~\cite{jenn,jeno}. We found that the computational time scales as $1.6^N$ in contrast to $2.7^N$ in the case of explicit enumeration, and all the densities of states for chain length of up to $N=42$ could be obtained within two days with a {\it single} CPU. 

However, the transfer matrix  method developed in this work is not meant to replace the explicit enumeration. Rather, the transfer matrix method is a tool which is complementary to the explicit enumeration. In the current form, the method can be used only for the square-lattice homopolymer, and cannot be applied to other lattice models such as three-dimensional polymer or HP protein models. Even in the context of the square-lattice homopolymer, only the quantities solely determined by the spatial distribution of the monomers, regardless of their positions {\it along the chain}, can be computed by the present method. The contact numbers considered in the current work, and other geometrical quantities such as radius of gyration~\cite{jenn,jeno}, are such examples. On the other hand, if we want to compute the average length of monomers connecting two monomers that are in spatial contact, the transfer matrix method cannot be used, because such nonlocal  information along the chain is not maintained in the building process of conformations. Another limitation is the requirement of memory resources, which is virtually null in the case of explicit enumeration. Therefore, in these situations where the transfer matrix method does not work, the explicit enumeration remains a valuable tool.

The tremendous amount of computational time required for the explicit enumeration of polymer chains have been overcome by parallelization~\cite{lee13,chen16}. We can expect the same for the transfer matrix algorithms. For the algorithm developed in the current work, the simplest method for parallelization would be to distribute the boxes to the nodes, because the enumerations of conformations spanning distinct boxes are independent tasks. The same idea has been used in the context of explicit enumeration~\cite{lee2}. The efficiency of this simple method is far from ideal, because the number of conformations vary greatly from box to box, and some of the nodes will keep working while the others have already finished the task. It is crucial to distribute the load evenly among the computational nodes to obtain maximal efficiency. In fact, an efficient parallel implementation of the transfer matrix method for enumerating self-avoiding polygon (SAP) has been developed where cut-line signatures rather than boxes are distributed among the nodes~\cite{para}. This idea is also similar to the recent efficient parallelization of the  explicit enumeration where the partial conformations rather than the boxes are distributed among the nodes~\cite{chen16}. Because the method developed in the current work is the extension of the method for SAW enumeration~\cite{jenn,jeno}, which is in turn the extension of the method for SAP~\cite{jenp,jenpo}, and thus share the common backbone structure, it is in principle straightforward to implement such parallelization. It is expected that the parallelization based on the distribution of cut-line signatures will also reduce the memory burden of each node, allowing us to compute the density of states for longer chains with ease.


\begin{acknowledgments}
This work was supported by the National Research Foundation of Korea, funded by the Ministry of Education, Science, and Technology (NRF-2014R1A1A2058188).
\end{acknowledgments}
\newpage

\begin{table}
\caption{Memory space required, $N_{\rm mem}$, as the functions of the chain length $N$. The factors contributing to $N_{\rm mem}$ are also shown.}
\scriptsize
\begin{tabular}{ c | r | r | r | r |}
\hline 
$N$ &  $K_{\rm max} +1$ & $N_c^{\rm max}$\footnote{The maximal number of combinations $(s_1, \cdots,s_{w+1},n)$, encountered during the progression of the algorithm.} &  $(K_{\rm max} +1) \cdot N_c^{\rm max}$  & $N_{\rm mem}$\footnote{The memory requirement in bytes}\\
\hline
3 & 1 & 0 & 0 &  0 \\
4 & 2 & 4 & 8 &  128 \\
5 & 2 & 11 & 22 &  352 \\
6 & 3 & 34 & 102 &  1\,632 \\
7 & 3 & 50 & 150 &  2\,400 \\
8 & 4 & 79 & 316 &  5\,056 \\
9 & 5 & 197 & 985 &  15\,760 \\
10 & 5 & 318 & 1\,590 &  25\,440 \\
11 & 6 & 416 & 2\,496 &  39\,936 \\
12 & 7 & 718 & 5\,026 &  80\,416 \\
13 & 7 & 1\,190 & 8\,330 &  133\,280 \\
14 & 8 & 1\,786 & 14\,288 &  228\,608 \\
15 & 9 & 2\,355 & 21\,195 &  339\,120 \\
16 & 10 & 3\,577 & 35\,770 &  572\,320 \\
17 & 10 & 5\,413 & 54\,130 &  866\,080 \\
18 & 11 & 7\,520 & 82\,720 &  1\,323\,520 \\
19 & 12 & 10\,819 & 129\,828 &  2\,077\,248 \\
20 & 13 & 16\,196 & 210\,548 &  3\,368\,768 \\
21 & 13 & 22\,768 & 295\,984 &  4\,735\,744 \\
22 & 14 & 32\,820 & 459\,480 &  7\,351\,680 \\
23 & 15 & 48\,165 & 722\,475 &  11\,559\,600 \\
24 & 16 & 68\,046 & 1\,088\,736 &  17\,419\,776 \\
25 & 17 & 99\,033 & 1\,683\,561 &  26\,936\,976 \\
26 & 17 & 143\,609 & 2\,441\,353 &  39\,061\,648 \\
27 & 18 & 206\,856 & 3\,723\,408 &  59\,574\,528 \\
28 & 19 & 296\,976 & 5\,642\,544 &  90\,280\,704 \\
29 & 20 & 428\,236 & 8\,564\,720 &  137\,035\,520 \\
30 & 21 & 626\,008 & 13\,146\,168 &  210\,338\,688 \\
31 & 21 & 890\,622 & 18\,703\,062 &  299\,248\,992 \\
32 & 22 & 1\,298\,532 & 28\,567\,704 &  457\,083\,264 \\
33 & 23 & 1\,886\,902 & 43\,398\,746 &  694\,379\,936 \\
34 & 24 & 2\,662\,054 & 63\,889\,296 &  1\,022\,228\,736 \\
35 & 25 & 3\,951\,437 & 98\,785\,925 &  1\,580\,574\,800 \\
36 & 26 & 5\,669\,758 & 147\,413\,708 &  2\,358\,619\,328 \\
37 & 26 & 8\,082\,368 & 210\,141\,568 &  3\,362\,265\,088 \\
38 & 27 & 11\,957\,233 & 322\,845\,291 &  5\,165\,524\,656 \\
39 & 28 & 17\,019\,325 & 476\,541\,100 &  7\,624\,657\,600 \\
40 & 29 & 24\,664\,128 & 715\,259\,712 &  11\,444\,155\,392 \\
41 & 30 & 36\,042\,443 & 1\,081\,273\,290 &  17\,300\,372\,640 \\
42 & 31 & 50\,797\,197 & 1\,574\,713\,107 &  25\,195\,409\,712 \\
\hline
\end{tabular}
\end{table}

\begin{table}
\caption{The densities of states for $N=41$ and $N=42$. The conformations related by rotation or reflection are counted only once.}
\scriptsize
\begin{tabular}{ c | r | r }
\hline 
K & \multicolumn{1}{c|}{N= 41} & \multicolumn{1}{c}{N= 42} \\
\hline
0 & 204\,215\,004\,596\,272 &  476\,389\,994\,800\,229 \\
1 & 885\,251\,445\,177\,512 &  2\,115\,847\,261\,636\,492 \\
2 & 2\,079\,694\,461\,161\,427 &  5\,084\,598\,808\,157\,791 \\
3 & 3\,464\,902\,826\,469\,576 &  8\,657\,251\,498\,020\,670 \\
4 & 4\,595\,250\,741\,229\,180 &  11\,720\,220\,074\,174\,806 \\
5 & 5\,156\,290\,616\,308\,466 &  13\,411\,555\,823\,963\,447 \\
6 & 5\,083\,784\,125\,308\,556 &  13\,473\,468\,289\,589\,989 \\
7 & 4\,512\,952\,758\,682\,396 &  12\,179\,635\,324\,640\,045 \\
8 & 3\,670\,155\,319\,845\,554 &  10\,081\,627\,576\,356\,447 \\
9 & 2\,768\,628\,794\,352\,198 &  7\,738\,288\,654\,944\,187 \\
10 & 1\,955\,769\,586\,807\,773 &  5\,560\,932\,203\,029\,585 \\
11 & 1\,302\,911\,413\,863\,672 &  3\,768\,601\,894\,558\,080 \\
12 & 823\,108\,147\,575\,701 &  2\,422\,174\,369\,581\,956 \\
13 & 495\,137\,600\,489\,206 &  1\,482\,830\,916\,877\,836 \\
14 & 284\,503\,636\,510\,917 &  867\,529\,474\,495\,897 \\
15 & 156\,477\,901\,312\,440 &  486\,206\,027\,920\,648 \\
16 & 82\,492\,824\,518\,467 &  261\,467\,916\,938\,889 \\
17 & 41\,702\,482\,928\,294 &  135\,033\,618\,385\,132 \\
18 & 20\,209\,523\,024\,356 &  66\,985\,846\,124\,393 \\
19 & 9\,372\,380\,538\,742 &  31\,891\,765\,478\,273 \\
20 & 4\,149\,991\,633\,601 &  14\,547\,520\,272\,987 \\
21 & 1\,746\,712\,880\,458 &  6\,339\,793\,465\,387 \\
22 & 693\,139\,648\,771 &  2\,627\,364\,285\,081 \\
23 & 257\,015\,560\,326 &  1\,027\,082\,613\,128 \\
24 & 87\,861\,707\,542 &  375\,429\,560\,638 \\
25 & 26\,600\,160\,006 &  125\,822\,825\,988 \\
26 & 6\,879\,377\,897 &  37\,442\,098\,467 \\
27 & 1\,268\,269\,356 &  9\,537\,009\,150 \\
28 & 95\,375\,740 &  1\,636\,035\,133 \\
29 & 744\,882 &  106\,244\,025 \\
30 & & 810\,017 \\
\hline
Total & 37\,599\,781\,156\,059\,284 & 100\,047\,629\,074\,894\,793 \\
\hline\end{tabular}
\end{table}

\newpage

\end{document}